\documentclass[runningheads]{llncs}
\pdfoutput=1
\usepackage{url}
\usepackage{graphicx}
\makeatletter
\newcommand{\@chapapp}{\relax}%
\makeatother

\usepackage{amssymb}
\usepackage{amsmath}
\usepackage[title]{appendix}
\usepackage{chngcntr}
\usepackage{listings}
\usepackage{xcolor}
\usepackage{xspace}
\newif\ifshowcomments
\showcommentsfalse

\ifshowcomments
\newcommand{\mynote}[2]{\fbox{\bfseries\sffamily\scriptsize{#1}}
 {\small$\blacktriangleright$\textsf{\emph{#2}}$\blacktriangleleft$}}
\else
\newcommand{\mynote}[2]{}
\fi

\definecolor{verylightgray}{rgb}{.97,.97,.97}

\lstdefinelanguage{Solidity}{
	keywords=[1]{anonymous, assembly, assert, balance, break, call, callcode, case, catch, class, constant, continue, contract, debugger, default, delegatecall, delete, do, else, emit, event, export, external, false, finally, for, function, gas, if, implements, import, in, indexed, instanceof, interface, internal, is, length, library, log0, log1, log2, log3, log4, memory, modifier, new, payable, pragma, private, protected, public, pure, push, require, return, returns, revert, selfdestruct, send, storage, struct, suicide, super, switch, then, this, throw, ttransfer, true, try, typeof, using, value, view, while, with, addmod, ecrecover, keccak256, mulmod, ripemd160, sha256, sha3}, 
	keywordstyle=[1]\color{blue}\bfseries,
	keywords=[2]{address, bool, byte, bytes, bytes1, bytes2, bytes3, bytes4, bytes5, bytes6, bytes7, bytes8, bytes9, bytes10, bytes11, bytes12, bytes13, bytes14, bytes15, bytes16, bytes17, bytes18, bytes19, bytes20, bytes21, bytes22, bytes23, bytes24, bytes25, bytes26, bytes27, bytes28, bytes29, bytes30, bytes31, bytes32, enum, int, int8, int16, int24, int32, int40, int48, int56, int64, int72, int80, int88, int96, int104, int112, int120, int128, int136, int144, int152, int160, int168, int176, int184, int192, int200, int208, int216, int224, int232, int240, int248, int256, mapping, string, uint, uint8, uint16, uint24, uint32, uint40, uint48, uint56, uint64, uint72, uint80, uint88, uint96, uint104, uint112, uint120, uint128, uint136, uint144, uint152, uint160, uint168, uint176, uint184, uint192, uint200, uint208, uint216, uint224, uint232, uint240, uint248, uint256, var, void, ether, finney, szabo, wei, days, hours, minutes, seconds, weeks, years},	
	keywordstyle=[2]\color{teal}\bfseries,
	keywords=[3]{block, blockhash, coinbase, difficulty, gaslimit, number, timestamp, msg, data, gas, sender, sig, value, now, tx, gasprice, origin},	
	keywordstyle=[3]\color{violet}\bfseries,
	identifierstyle=\color{black},
	sensitive=false,
	comment=[l]{//},
	morecomment=[s]{/*}{*/},
	commentstyle=\color{gray}\ttfamily,
	stringstyle=\color{red}\ttfamily,
	morestring=[b]',
	morestring=[b]"
}

\DeclareMathOperator{\size}{size}
\DeclareMathOperator{\gas}{gas}
\DeclareMathOperator{\bctime}{time}
\DeclareMathOperator{\txblock}{block}
\newcommand{\BlockMaxSize}{S^{\mathrm{max}}}
\newcommand{\GasLimit}{G}
\newcommand{\TransactionSet}{\mathcal{SC}}
\newcommand{\TransactionType}{TX}
\newcommand{\Transaction}{tx}
\newcommand{\Block}{b}
\newcommand{\TransactionLatency}{L}
\newcommand{\ConsensusLatency}{L_{\mathrm{C}}}
\newcommand{\BlockInclLatency}{L_{\mathrm{B}}}
\newcommand{\HeaderSize}{s_{\mathrm{H}}}
\newcommand{\TransferSize}{s_{\mathrm{T}}}
\newcommand{\GenesisBlockSize}{s_{\mathrm{g}}}
\newcommand{\TransferGas}{g_{\mathrm{T}}}
\newcommand{\BlockPeriod}{T}
\newcommand{\HeadersOverhead}{\operatorname{OH}}
\newcommand{\GrowthRate}{\operatorname{GR}}

\newcommand{\arch}{B-CoC\xspace}

\newcommand{\ie}{i.e.,\xspace}
\newcommand{\eg}{e.g.,\xspace}

\begin{document}
\counterwithin{lstlisting}{section}
%
\title{\arch: A Blockchain-based Chain of Custody for Evidences Management in Digital Forensics}
\titlerunning{\arch for Evidences Management in Digital Forensics}
%
\author{Silvia Bonomi\inst{1, 2} \and
Marco Casini\inst{2} \and
Claudio Ciccotelli\inst{1, 2}}
\authorrunning{S. Bonomi et al.}
%
\institute{Research Center of Cyber Intelligence and Information Security (CIS) \and
Department of Computer, Control, and Management Engineering ``A. Ruberti'',\\
Sapienza Universit\`{a} di Roma, Via Ariosto 25, 00185 Roma, Italy\\
\email{\{bonomi,ciccotelli\}@diag.uniroma1.it, casini.1724011@studenti.uniroma1.it}}
\maketitle              
\begin{abstract}
One of the main issues in digital forensics is the management of evidences. 
From the time of evidence collection until the time of their exploitation in a legal court, evidences may be accessed by multiple parties involved in the investigation that take temporary their ownership. This process, called \emph{Chain of Custody} (CoC), must ensure that evidences are not altered during the investigation, despite multiple entities owned them, in order to be admissible in a legal court.
Currently digital evidences CoC is managed entirely manually with entities involved in the chain required to fill in documents accompanying the evidence.
In this paper, we propose a Blockchain-based Chain of Custody (B-CoC) to dematerialize the CoC process guaranteeing auditable integrity of the collected evidences and traceability of owners.
We developed a prototype of B-CoC based on Ethereum and we evaluated its performance.

\end{abstract}
%
%
%


\section{Introduction}\label{sec:introduction}

%
%

One of the main issues in digital forensics is the management of evidences. 
From the time of evidence collection until the time of their exploitation in a legal court, evidences may be accessed by multiple parties involved in the investigation that take temporarily their ownership. 
The \textit{Chain of Custody} is the process of validating how any kind of evidence has been gathered, tracked and protected on its way to a court of law.
Chain of Custody (CoC) is not a mandatory step in forensic analysis. However, it is extensively used as evidences, to be acceptable in a court or in legal procedures, must be proved to be not altered during investigations.
Thus, a good CoC process should use a standard for dealing and handling evidences (digital or not), regardless of whether the evidence will be used in a trial or not. 

The main requirements of a CoC process are:

\begin{itemize}
\item {\bf Integrity}: the evidence has not been altered or corrupted during the transferring.
\item {\bf Traceability}: the evidence must be traced from the time of its collection until it is destroyed.
\item {\bf Authentication}: all the entities interacting with an evidence must provide an irrefutable sign as a recognizable proof of their identity.
\item {\bf Verifiability}: the whole process must be verifiable from every entity involved in the process.
\item {\bf Security - Tampering proof}: Changeovers of an evidence cannot be altered or corrupted.
\end{itemize}

Currently, CoC process requirements are met by employing a physical handover of evidences where, at each step, documents are filled in and signed in front of officers.
In this paper, we take a step toward the dematerialisation of this process by proposing a Blockchain-based architecture for CoC of digital evidences.

Leveraging on the features offered by blockchain technologies, we defined an architecture able to support the CoC process. 
In particular, we set up a private permissioned blockchain and we implemented a smart contract to keep track of the ownership changes during the evidence lifecycle.
We implemented our prototype on an Ethereum~\cite{wood2014ethereum} private network and we evaluated the impact of the system configuration parameters on performance.


\section{Background}\label{sec:relatedwork}


\subsection{Blockchain technology}

The blockchain technology implements a decentralized fully replicated append-only ledger in a peer-to-peer network, originally employed for the Bitcoin cryptocurrency~\cite{nakamoto2008bitcoin}.
All participating nodes maintain a full local copy of the blockchain.
The blockchain consists of a sequence of blocks containing the transactions of the ledger. 
Transactions inside blocks are sorted chronologically and each block contains a cryptographic hash of the previous block in the chain.
Nodes create new blocks as they receives transactions, which are broadcast in the network.
Once a block is complete, they start the consensus process to convince other nodes to include it in the blockchain.
In the original blockchain technology employed in Bitcoin the consensus process is based on \emph{Proof-of-Work} (PoW)~\cite{nakamoto2008bitcoin}.
With PoW nodes compete with each other in confirming transactions and creating new blocks by solving a mathematical puzzle.
While solving a block is a computational intensive task, verifying its validity is easy.
To incentivize such mechanism, solving a block also results in mining a certain amount of bitcoins, which is the reward for block creators (usually referred to as \emph{miners}).
Sometimes, more than one miner may generate a valid block thus creating forks in the chain.
Forks are solved by accepting only the longest branch as the valid continuation of the chain (thus eliminating forks eventually).
The main advantage of PoW, over traditional consensus algorithms, is that an attacker would have to control the majority of the computational power of the network, rather than the majority of the nodes, which is considered more difficult and virtually impossible in public large-scale networks.

The main criticism to PoW is its huge demand of energy, which also prevents its applicability in certain contexts.
This has led to the investigation of alternative forms of consensus for the blockchain, such as \emph{Proof-of-Stake}~\cite{DBLP:conf/fc/BentovGM16}.
With PoS, a set of nodes, called \emph{validators}, take turns proposing new blocks and voting on them.
Validators put a stake in the network (\eg a given amount of cryptocurrency) and are incentivized to act honestly so as not to lose the stake.
Indeed, the blockchain keeps track of the set of validators, which are ousted if they behave maliciously (thus losing their stake).

A specific type of PoS is \emph{Proof-of-Authority} (PoA) in which individual's identity (rather than cryptocurrency) is at stake.
With PoA validators must have been preventively authorized and their identities are known.
Thus, acting maliciously results in losing personal reputation and ultimately in being expelled from the validator set.

While PoW is particularly suited for \emph{public} networks, both PoS and PoA may be suitable for \emph{private} networks (where PoW would probably fail short as it would be much easier to control the majority of the computational power). 
Moreover, PoW and PoS can be used in \emph{permissionless} networks, that is, networks where nodes can freely join the network without previous authorization (\eg as in Bitcoin and Ethereum).
PoA, on the other hand, is typically employed in \emph{permissioned} blockchain networks, that is, networks in which nodes cannot freely join and become validators, but rather they have to be preventively authorized.

\subsection{Ethereum and Smart Contracts}\label{sec:ethereum}

Etherium~\cite{wood2014ethereum} can be seen as a decentralized virtual machine based on the blockchain technology.
The Ethereum Virtual Machine (EVM) runs programs, referred to as \emph{smart contracts}, whose state is stored in the Ethereum blockchain.
Every node execute a local EVM.
When an account wants to execute a function of a smart contract, it issues a transaction which is broadcast to the network.
Each node executes the transaction on its local EVM and stores it, along with the new computed state, in the blockchain.



In Ethereum each EVM instruction consumes a virtual resource referred to as \emph{gas}. 
Gas can be seen as the fuel of the EVM and is employed to incentivize miners to execute transactions and include them in the blockchain.
Indeed, for each transaction, miners are rewarded by the issuer with the payment of fees proportional to the total amount of gas ``consumed'' to execute that transaction.
%
To prevent mined blocks from becoming too large, which may severely impact block propagation and processing latency, each block has a \emph{block gas limit}, which is the maximum amount of gas all transactions included in the block are allowed to consume. 
Thus, an issued transaction may not be included in the current block by a miner because it would exceed the block gas limit. 
In such case, the issued transaction would have to wait until the next block creation.

The public Ethereum blockchain (often referred to simply as ``Ethereum'') is a public permissionless networks which adopts PoW as consensus algorithm (even though it is planned to switch to PoS in the future).
However, all major Ethereum implementations~\cite{geth,parity} allow to configure many aspects of the protocol, such as the actual consensus algorithms employed, and allow to build custom public/private permissionless/permissioned blockchain networks.

\subsection{Istanbul BFT consensus protocol}\label{sec:ibft}

Istanbul Byzantine Fault Tolerance (IBFT)~\cite{ibft} is an adaptation of the Practical Byzantine Fault Tolerance (PBFT)~\cite{Castro:1999:PBF:296806.296824} algorithm to serve as a PoA consensus algorithm for the Ethereum protocol.
IBFT can tolerate at most $f$ faulty validators out of a total of $n = 3f + 1$ validators.
The IBFT algorithm proceeds in rounds with a new block created every $\BlockPeriod$ seconds, where the \emph{block period} $\BlockPeriod$ is a constant configuration parameter.
In each round one of the validators is elected as the \emph{proposer}.
The proposer creates the new block and broadcasts it to all validators with a \emph{pre-prepare} message.
Upon receiving pre-prepare messages, validators enter the \emph{pre-prepared} phase and broadcast \emph{prepare} messages.
This, ensures that validators are aligned to the same round and block.
Upon receiving $2f + 1$ prepare messages, validators enter the \emph{prepared} phase and broadcast \emph{commit} messages to inform other validators that they accept the proposed block.
Finally, upon receiving $2f + 1$ commit messages, validators enter the \emph{committed} phase and insert the block in the blockchain.


\section{System Model}\label{sec:model}

\noindent{\bf CoC model.} A digital evidence (or electronic evidence) is any probative information stored or transmitted in digital form that a party may use in a trial to a court case. Digital evidences are collected by authorised parties (usually police officers) that become their temporary (first) owners.

For the sake of presentation and without loss of generality, in the following we will consider a single digital evidence $d\_ev$ collected by an authorised entity $e_0$ that holds its ownership.
During investigations, several authorised entities (\eg police offices, lawyers, judges, magistrates, etc.) may need to access, acquire and/or own temporarily $d\_ev$. 
The set of authorised entities that can interact with $d\_ev$ is denoted with $A_{d\_ev}$.
%
Each authorised entity has a unique identifier known to all and he/she owns credentials that allows him/her to be authenticated and take actions in the CoC process.

At each time $t$, $d\_ev$ can have just one \emph{owner} and the owner must belong to $A_{d\_ev}$. If an authorised entity $e_i$ needs to acquire and own $d\_ev$, the current owner needs to issue a \emph{transfer} request towards $e_i$.
The change of ownership happens if and only if $e_i \in A_{d\_ev}$ and the transfer record is written permanently in the \emph{evidence log}.\\ 


\noindent{\bf Network Model.} The system is composed by a set of processes $p_1, p_2, \dots, p_n$, one for each authorised entity in $A_{d\_ev}$. Each process $p_i$ has a pair of private-public keys that it uses to authenticate itself and to sign messages.
Processes are connected trough a peer-to-peer network (authenticated perfect links).


\section{B-CoC Architecture}\label{sec:architecture}

%

The Blockchain-based Chain of Custody (\arch) architecture proposed in this paper is based on a \textit{private} and \textit{permissioned} blockchain.
This choice has been driven by the \emph{authentication} requirement of the CoC process that does not allow unauthorised and untrusted parties to manage digital evidences and thus to be in the network. 

\begin{figure}[t]\centering
\includegraphics[trim={0 1.8cm 0 2.5cm},clip,width=\linewidth]{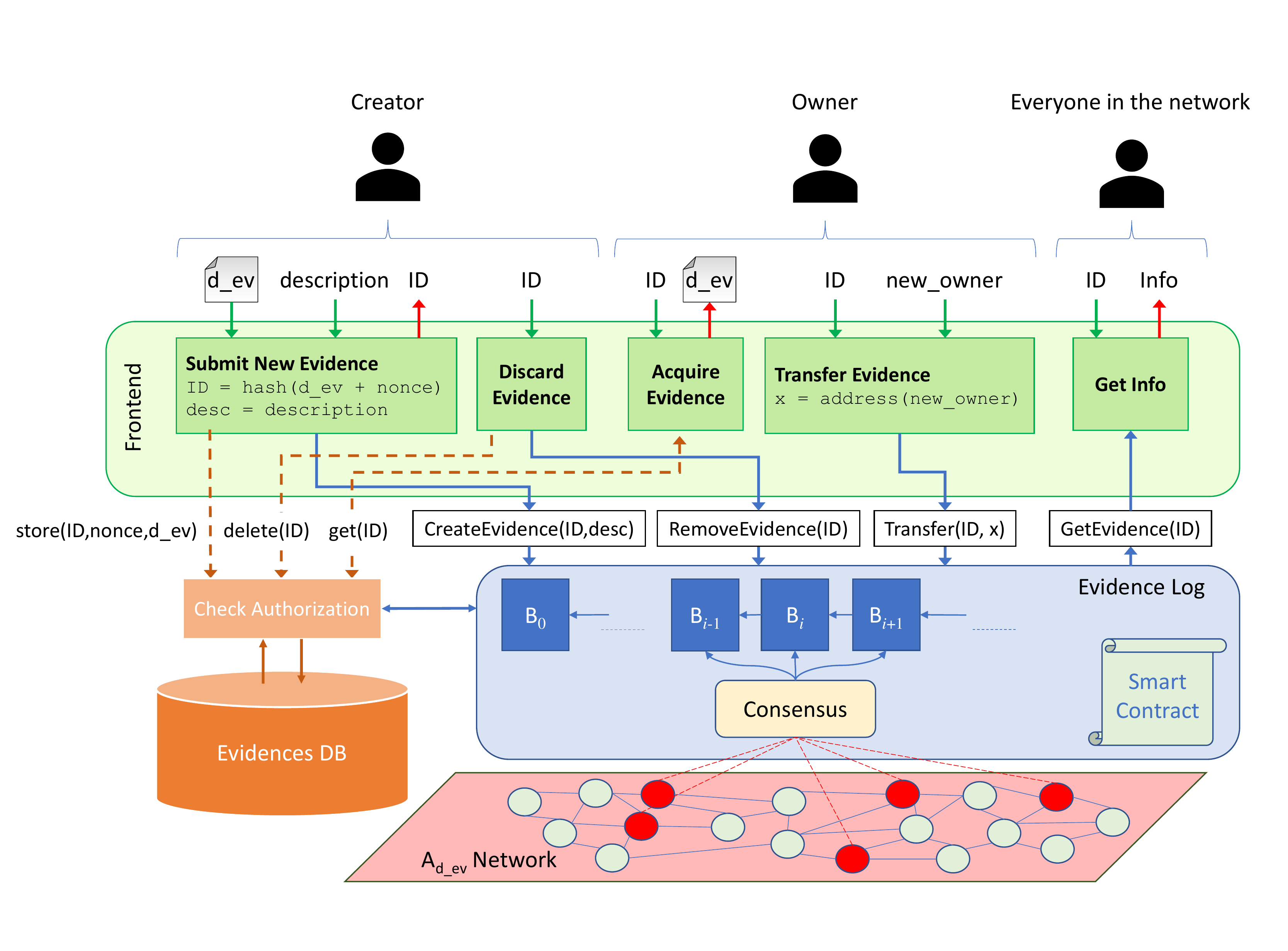}
\caption{\arch architecture.}\label{fig:architecture}
\end{figure}

As shown in Figure~\ref{fig:architecture}, \arch is composed mainly of three components: (i)~the \emph{Evidences DB}, (ii)~the \emph{Evidence Log} and the (iii)~\emph{Frontend} interface.
The Evidence DB is an ordinary database and/or file repository where we store the actual digital evidences, while CoC related data are stored in the Evidence Log, which is implemented through the blockchain technology.
The reason for this separation is twofold.
First of all, evidences can be too large to be efficiently stored in the blockchain (for example, an evidence may be a bit-by-bit copy of a storage device of several TBs of capacity). 
Secondly, and most importantly, if evidences were stored in the blockchain, every node in the blockchain network would have access to them, while only authorized nodes should be allowed to acquire an evidence.
Therefore, we store in the blockchain only the information regarding the CoC process and an hash of the evidence which allows to verify evidences integrity during acquisition.\\

\noindent\textbf{Evidence DB.} The Evidences DB is an ordinary database and/or file repository where the original digital evidence is stored along with an identifier \texttt{ID}, obtained as the hash of the evidence and a nonce (to guarantee uniqueness of \texttt{ID}s).
This database is distributed and is managed by trusted entities (\eg law court officers).
Moreover, each access is executed only if the requesting entity is authorized to perform such access according to its role.\\

\noindent\textbf{Evidence Log.}
The Evidence Log is implemented trough the blockchain technology and stores, for each evidence, its \texttt{ID}, a description, the identity of the submitter (which we call \emph{creator}) and the complete history of owners up to the current one, including the time at which changes of ownership occurred.
Note that while the evidence itself is not stored in the blockchain, the \texttt{ID} allows to verify that the evidence has not been tampered with, provided that a robust cryptographic hash function is used to generate it.

The evidence log is implemented on top of a peer-to-peer network composed by all authorised entities.
Such network can be decomposed in two sets of nodes:
\begin{itemize}
\item \textit{Validator nodes}: they have mainly the following functionalities: (i) storing a copy of the blockchain, (ii) validating transactions and (iii) create, propose and add blocks to the chain (\ie participate to the consensus protocol). 
This is the set of nodes that must be preventively authorized with the role of validators in the permissioned blockchain.
%
\item \textit{Lightweight nodes}: they can be seen as clients of the chain since they simply issue transactions and need to rely on validators for adding and validating their transactions.
\end{itemize}
Taking Italy as a use case, each validator may correspond to the main coordinator of the court of one of the 20 regional capitals.
Lightweight nodes, instead, would represent all the other involved parties such as police departments, forensic investigators, forensic consultants and so on.  
The Evidence Log runs a smart contract which exposes four primitives (see Figure~\ref{fig:architecture}):
\begin{itemize}
\item \texttt{CreateEvidence(ID, description)}: stores a new evidence entry in the blockchain with the specified \texttt{ID} and \texttt{description}, setting the submitter identity as the \emph{creator} and current \emph{owner} of the evidence.
\item \texttt{Transfer(ID, newowner)}: transfers the ownership of an evidence (registering the handover). It fails if the issuer is not the current owner.
\item \texttt{RemoveEvidence(ID)}: removes an evidence entry. It fails if the issuer is not the creator.
\item \texttt{GetEvidence(ID)}: returns the information in the evidence entry. Namely, the \texttt{ID}, description, creator and all owners with the time of each change of ownership. 
\end{itemize}
Implementation details of the Evidence Log and the smart contract are discussed in Section~\ref{sec:implementation}.\\

\noindent\textbf{Frontend Interface.}
The frontend represents the interface between \arch and its users.
A local instance runs on each node and interacts with the Evidences DB and the Evidence Log (through a local blockchain node).
When an authorized user submits a new digital evidence $d\_ev$ to the system, he/she takes the role of \emph{creator} of $d\_ev$ (see Figure~\ref{fig:architecture}).
The frontend generates the \texttt{ID} for $d\_ev$ using a nonce $n$, stores $(\texttt{ID}, n, d\_ev)$ in the Evidence DB and issues the \texttt{CreateEvidence()} transaction in the Evidence Log.
As already discussed the submitter is also registered as the first owner in the blockchain.

The creator of an evidence $d\_ev$ can request to discard it from the system (\eg because it is no more legally valid). 
If he/she is authorized to do so, the corresponding entry is removed from the Evidence Log by issuing the \texttt{RemoveEvidence()} transaction. 
If the transaction succeeds, the corresponding evidence is deleted from the Evidence DB.

When a user wants to acquire an evidence $d\_ev$, the Frontend sends a request to the Evidences DB which will serve the request only if the user is the current owner of $d\_ev$.  This check is performed by interacting with the Evidence Log.

The change of ownership of an evidence $d\_ev$ is performed by issuing a \texttt{Transfer()} transaction specifying the new owner.

Finally, every user in the \arch network can query the Evidence Log to get the entry of an evidence (which contains all relevant information except the evidence itself).
This is performed by simply issuing the \texttt{GetEvidence()}.




\section{Evidence Log Implementation}\label{sec:implementation}

As described in Section~\ref{sec:architecture}, \arch Evidence Log is designed as a private and permissioned blockchain.
The blockchain infrastructure is implemented through \emph{Geth}~\cite{geth} a popular implementation of a full Ethereum node.
Geth allows to setup a private network and configure all aspects of the blockchain and the consensus protocol employed.
Given the design of a private permissioned blockchain we adopt a PoA-based consensus.
Namely, the IBFT consensus protocol described in Section~\ref{sec:ibft}.
On top of this blockchain infrastructure, we run a smart contract implementing the CoC process.
The implementation of \arch Evidence Log involves three steps: (i) the initialization of the private blockchain, (ii) the creation of the private network and (iii) the creation and deployment of the smart contract.

\subsection{Private chain initialization}
The setup of a new blockchain involves the creation of its \emph{genesis} block.
This is the first block of a blockchain and contains the initial parameters.
The only configuration parameters that are of interest for the purposes of the following discussion are:
\begin{itemize}
	\item \emph{Block Period} $T$: the block period of the IBFT consensus algorithm (see section~\ref{sec:ibft}); 
	\item \emph{Block Gas Limit} $G$: Maximum amount of gas transactions in a block are allowed to consume (see section~\ref{sec:ethereum});
	\item \emph{Validators}: The Ethereum addresses of the pre-authorized validators.
\end{itemize}
The genesis block is used to initialize each node of the network.

\subsection{Private network setup}

First of all, to build the private peer-to-peer network we need to setup the peer discovery service to allow new nodes to enter the network and know other nodes. 
This is accomplished with the \texttt{bootnode} tool (of the Geth tools suite).
This tool allows to run special nodes (with known IP addresses) that validators and lightweight nodes will contact when first started to exchange peer information.

Validators and lightweight nodes are Geth nodes.
First, we configure the set of validators (which is fixed and known in advance) with the genesis block and we run them through the \texttt{geth} command (of the Geth tools suite).
Validators are created once at the beginning and they never leave the network, unless they act maliciously and are expelled.
Lightweight nodes, instead, can be created and join/leave the network at any time.
They are created with the \texttt{geth} tool as well, but their addresses are not included in the genesis block.

\subsection{Smart contract implementation}

The smart contract has been implemented through the Solidity contract-oriented programming language~\cite{solidity}.
Due to space constraints, the code of the smart contract is reported in the Appendix. 
The smart contract manages entries associated to digital evidences (\ie the entries of the Evidence Log).
Each \texttt{Evidence} entry (lines~\ref{line:struct-start}-\ref{line:struct-end}) consists of the \texttt{ID}, the Ethereum address of the creator, the address of the owner, a string field to store the description of the evidence and two arrays \texttt{taddr} and \texttt{ttime} that store, respectively, the evidence handovers and the times at which they occurred.
These arrays are chronologically sorted from the creator to the current owner.
All evidence items are stored in a map indexed by evidence \texttt{ID}s (line~\ref{line:map}).
The smart contract has a total of four functions implementing the primitives of the Evidence Log described in Section~\ref{sec:architecture}.
The \texttt{CreateEvidence(ID, description)} function creates a new \texttt{Evidence} entry with the specified \texttt{ID} and \texttt{description}, and the address of the related transaction sender as the creator and current owner of the evidence (line~\ref{line:create-evidence}).
The \texttt{Transfer(ID, newowner)} function transfers the ownership of the evidence identified by \texttt{ID} to the entity identified by the address \texttt{newowner} (line~\ref{line:transfer}). 
Note that only the current owner of an evidence can transfer ownership (\texttt{OnlyOwner} modifier).
The \texttt{RemoveEvidence(ID)} function removes an evidence from the map of evidences (line~\ref{line:remove-evidence}). 
No further operations can be performed on a removed evidence. 
Note that only the creator of an evidence can remove the evidence (\texttt{OnlyCreator} modifier).
The \texttt{GetEvidence(ID)} function returns all fields of an evidence entry (line~\ref{line:get-evidence}).

Note that while calling the first three functions results in issuing transactions to the blockchain that modify the state of the smart contract, the \texttt{GetEvidence} function only returns an entry and does not modify the state.
In the context of the Solidity language these are called \emph{constant} functions or \emph{views}.
Calling views does not result in the issuing of transactions, but rather they are executed locally by the node's local EVM.


\section{Evaluation}\label{sec:evaluation}

In this section we evaluate how the parameters of \arch, namely the block period $\BlockPeriod$ and the block gas limit $\GasLimit$, affect its performance. Section~\ref{sec:tx_latency} reports an analysis of the transaction latency, Section~\ref{sec:space_overhead} evaluates the space overhead due to block headers and Section~\ref{sec:growth_rate} discusses the growth rate of the blockchain.\\

\begin{table}[t]
\centering
\caption{Size and gas used by each transaction.}
\label{tab:size_gas}
\begin{tabular}{|l|r|r|}
\hline
 $\TransactionType$ & $\size(\TransactionType)$ (bytes) & $\gas(\TransactionType)$ (units) \\
\hline
\hline
\texttt{CreateEvidence(0)}    &  207 &  170207 \\
\hline
$\cdots$             & $\cdots$ & $\cdots$ \\
\hline
\texttt{CreateEvidence(1024)} & 1233 &  897367 \\
\hline
\texttt{Transfer()}           &  174 &   80502 \\
\hline
\texttt{RemoveEvidence()}     &  142 &  236478 \\
\hline
\end{tabular}
\end{table}

\noindent{\bf Notation.} In the following sections we will use the notation $\TransactionType$ to refer to a \emph{transaction type} (\ie a non-constant function of a smart contract) and $\Transaction$ to refer to an \emph{execution of a transaction}. 
For example, $\TransactionType$ may refer to the transaction type \texttt{Transfer()}, while $\Transaction$ may refer to an actual execution of a \texttt{Transfer()} of an evidence.
We will use $\gas(\Transaction)$ and $\size(\Transaction)$ to indicate, respectively, the gas consumed by the execution of a transaction $\Transaction$ and the size (in bytes) of $\Transaction$ when included in a block.
Note that, in general, both $\gas(\Transaction)$ and $\size(\Transaction)$ depend on the particular execution of $\Transaction$.
In practice, for our smart contract, transaction types \texttt{Transfer()} and \texttt{RemoveEvidence()} have constant size and gas used, while for \texttt{CreateEvidence()} such parameters depend exclusively on the length $\ell$ of the \texttt{description} parameter.
Thus, for ease of presentation we will consider a different transaction type \texttt{CreateEvidence($\ell$)} for each value of $\ell$.
Since we limit the length of the \texttt{description} parameter to 1024 characters, we consider 1025 different transaction types ($\ell = 0,\dots,1025$). 
Thus, each transaction type has constant size and constant consumed gas and, therefore, we will consider $\size(\TransactionType) = \size(\Transaction)$ where $\Transaction$ is an execution of $\TransactionType$ and use the two members of the equation interchangeably, as well as $\gas(\TransactionType) = \gas(\Transaction)$.
We will refer to $\TransactionSet = \lbrace \TransactionType_1, \dots, \TransactionType_n \rbrace$ as the set of transaction types of the smart contract. 
Table~\ref{tab:size_gas} reports the size and gas of the transaction types in our smart contract.
Due to space constraints Table~\ref{tab:size_gas} only shows \texttt{CreateEvidence($\ell$)} for $\ell=0$ and $\ell=1024$, but the size and gas used by such transaction types increase with $\ell$.

\subsection{Transaction latency}\label{sec:tx_latency}

The transaction latency $\TransactionLatency(tx) = \BlockInclLatency(\Transaction) + \ConsensusLatency(\Block)$ is the time elapsed from the issue of the transaction to its inclusion in the blockchain.
It is the sum of the \emph{block inclusion latency} $\BlockInclLatency(\Transaction)$, that is the time required by $\Transaction$ to be included in a block $\Block$ of the current proposer, and the \emph{consensus latency} $\ConsensusLatency(\Block)$, which is the time required to reach consensus on block $b$ and include it in the blockchain:
%
%
%
In the next two sections we will analyze these two terms separately.

\subsubsection{Block inclusion latency}\label{sec:block_incl_latency}

The block inclusion latency $\BlockInclLatency(\Transaction)$ is the time required for a transaction $\Transaction$ to be included in a block.
Indeed, whenever a new transaction is issued it may not \emph{fit} in the block of the current proposer due to the block gas limit $\GasLimit$.
In such case, the transaction is reissued in the next block period.

More formally, let $\txblock(\Transaction)$ be the block in which, eventually, transaction $\Transaction$ is included, and $\bctime(\Transaction)$, $\bctime(\Block)$ be, respectively, the time at which $\Transaction$ is issued and the time at which a block $\Block$ was created (i.e., the beginning of $b$'s block period), then:
\begin{equation}\label{eqn:block_incl_latency}
	\BlockInclLatency(\Transaction) = \bctime(\txblock(\Transaction)) + \BlockPeriod - \bctime(\Transaction)
\end{equation}
where $\BlockPeriod$ is the block period.

The block inclusion latency is affected by the block period parameter $\BlockPeriod$, the gas limit $\GasLimit$ and the workload, i.e., the rate of transactions issued to the system in the unit of time.
Suppose that we are able to precisely characterize the workload the system is subject to and to set $\GasLimit$ such that every issued transaction is included in the block of the current proposer. 
In such ideal conditions, ${\BlockInclLatency(\Transaction) \in [0, \BlockPeriod]}$. 
That is, the maximum block inclusion latency is the block period~$\BlockPeriod$.
Clearly, setting $\GasLimit = \infty$ would meet the ideal conditions for every possible workload volume, but on the other side would negatively impact consensus latency, as blocks could increase indefinitely (see next section).
Thus, we would like to set $\GasLimit$ as small as possible (to reduce consensus latency), but large enough so that (at least on average) every transaction is included in the block of the current proposer.
Thus the ideal value of $\GasLimit$ depends on the workload.
However, rather than the number of transactions per seconds, it depends on the \emph{gas rate}, \ie the amount of gas consumed by the transactions issued in a given block period.
Indeed, the minimum value of $\GasLimit$ that guarantees the ideal conditions for the block inclusion latency is the maximum gas rate.

\begin{figure}[t]\centering
\includegraphics[width=0.6\linewidth]{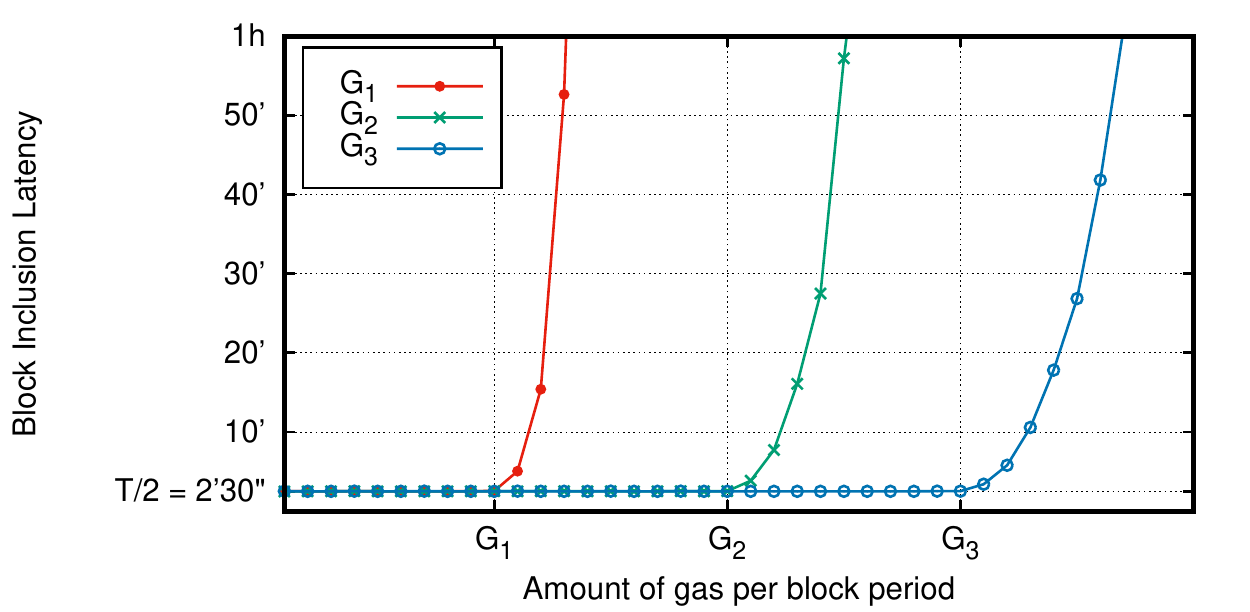}
\caption{Mean block inclusion latency varying the \emph{gas rate}, i.e., the amount of gas consumed by transactions in a block period.}\label{fig:block_incl_latency}
\end{figure}

Figure~\ref{fig:block_incl_latency} shows the results of three experiments that confirm the previous claim.
In each experiment we set a different value of the block gas limit ($\GasLimit_1$, $\GasLimit_2$ $\GasLimit_3$) and we progressively increased the gas rate from the start to the end of the experiment.
From the figure we can clearly see that, in each experiment, when the gas rate is less than or equal to the gas limit the average block inclusion latency is approximately equal to the expected value of $\BlockPeriod / 2$ (because transactions are issued uniformly distributed in each block period) while the maximum latency is $\BlockPeriod$ (not shown in the figure).
However, as soon as the gas rate exceeds the block gas limit the average block inclusion latency starts increasing indefinitely as expected.

This analysis provides a lower bound for the value of the block gas limit (\ie the maximum gas rate), that allows to minimize the maximum block inclusion latency to $\BlockPeriod$.
However, determining such value may be difficult.
Appendix~\ref{sec:discussion-params} reports a more general and detailed discussion on setting the parameters of \arch.

\subsubsection{Consensus latency}\label{sec:consensus_latency}

Given the consensus protocol described in section~\ref{sec:ibft}, the consensus latency, i.e., the time required to propagate a block $\Block$ between the validators and reach consensus, can be approximated by the formula
$\ConsensusLatency(\Block) \approx \frac{s_\mathrm{PP}(\Block) + s_\mathrm{P} + s_\mathrm{C}}{R}$, where $s_\mathrm{PP}(\Block)$ is the size of the \emph{pre-prepare} message, $s_\mathrm{P}$ is the size of the \emph{prepare} message, $s_\mathrm{C}$ is the size of the \emph{commit} message and $R$ is the bandwidth of the slowest communication channel between two validators nodes (bytes/sec).
While $s_\mathrm{P}$ and $s_\mathrm{C}$ are constant, the pre-prepare message piggybacks the block $\Block$ and thus $s_\mathrm{PP}(\Block)$ depends on $\size(\Block)$.
Since $R$ is typically a constant that depends on the infrastructure connecting the validator nodes, the only factor that we can adjust to control the latency is the size of a block.

The size of a block is the sum of the size of the transactions in it plus the size of the block header $\HeaderSize$ (which is constant). 
In our prototype implementation of \arch, $\HeaderSize = 1909$ bytes.
The actual number and type of transactions in a block depends on many factors, including the block period $\BlockPeriod$, the block gas limit $\GasLimit$ and ultimately the particular set of transactions sent during a given time period. 
Thus, in general, different blocks have different sizes.
However, we can control the maximum block size $\BlockMaxSize$, and thus the maximum consensus latency $\ConsensusLatency^{\mathrm{max}}$, by adjusting the block gas limit $\GasLimit$.

For a given value of $\GasLimit$, the maximum block size $\BlockMaxSize$ can be computed by solving the following optimization problem:

\begin{problem}[UKP]\label{prob:ukp}
\[
\begin{aligned}
& \operatorname{maximize} & & \sum\limits_{\TransactionType_i \in \TransactionSet} \size(\TransactionType_i) \cdot x_{i} \\
& \text{subject to} & & \sum\limits_{\TransactionType_i \in \TransactionSet} \gas(\TransactionType_i) \cdot x_i \leq \GasLimit \\
& & & x_i \in \mathbb{N}, \quad i = 1, \dots, n
\end{aligned}
\]
\end{problem}
where $x_i$ is the number of times a transaction of type $\TransactionType_i$ appears in the block of maximum size. 
The optimal solution $\lbrace x^*_1, \dots, x^*_n \rbrace$ leads to the maximum block size:
\[
	\BlockMaxSize = \HeaderSize + \operatorname{OPT_{UKP}}(\GasLimit) = \HeaderSize + \sum\limits_{\TransactionType_i \in \TransactionSet} \size(\TransactionType_i) \cdot x^*_{i}
\]
%
%
Problem~\ref{prob:ukp} is an instance of the well-known unbounded knapsack problem~\cite{POIRRIEZ2009110}, where transaction types correspond to the items to fit in the knapsack, while transactions' size and consumed gas correspond, respectively, to items' value and weight. 
The block gas limit parameter $\GasLimit$ corresponds to the knapsack maximum weight. 

While the general unbounded knapsack problem is NP-hard (with time complexity $O(n\GasLimit)$), this particular instance turns out to be trivial.
Indeed, it is easy to see that \texttt{Transfer()} \emph{dominates} all other transaction types~\cite{POIRRIEZ2009110}. 
That is, given any block containing at least a transaction $\Transaction$ of type in $\TransactionSet \setminus \lbrace \texttt{Transfer()} \rbrace$, we can always replace $\Transaction$ with a sufficient number of \texttt{Transfer()} so as to obtain a better solution to Problem~\ref{prob:ukp}.
For example, we can always replace a transaction of type \texttt{RemoveEvidence()} with a single \texttt{Transfer()} and obtain a solution that consumes less gas but have larger size. 
The same occurs if we replace \texttt{CreateEvidence(0)} with $2$ \texttt{Transfer()}, or \texttt{CreateEvidence(1024)} with at least $8$ \texttt{Transfer()}.
This implies that the optimal solution of this instance of the unbounded knapsack problem corresponds to a block consisting of only \texttt{Transfer()} transactions. 
Therefore, let $\TransferGas = \gas(\texttt{Transfer()})$, $\TransferSize = \size(\texttt{Transfer()})$, the solution is simply given by:
\begin{equation}\label{eqn:block_max_size}
	\BlockMaxSize = \HeaderSize + \left\lfloor \frac{\GasLimit}{\TransferGas} \right\rfloor \cdot \TransferSize
\end{equation}
Note that $\BlockMaxSize$ cannot be an arbitrary integer, but only one such that $\BlockMaxSize = \HeaderSize + k \cdot \TransferSize$, $k \in \mathbb{N}$.
%
%
Once chosen the appropriate value of $\BlockMaxSize$ (one that allows to limit the maximum consensus latency to an acceptable bound), $\GasLimit$ can be set to any value such that:
\begin{equation}\label{eqn:gas_limit}
	\GasLimit = \frac{\BlockMaxSize - \HeaderSize}{\TransferSize} \cdot \TransferGas + r = k \cdot \TransferGas + r, \qquad r = 0, \dots, \TransferGas - 1
\end{equation}
%
%
%
%
Equations~\ref{eqn:block_max_size} and~\ref{eqn:gas_limit} allows to determine an upper-bound for $\GasLimit$ so as to bound the maximum consensus latency $\ConsensusLatency^\mathrm{max}$. 
A more general discussion on how to properly set \arch's parameters is reported in Appendix~\ref{sec:discussion-params}.

\subsection{Block headers overhead}\label{sec:space_overhead}



As discussed in section~\ref{sec:block_incl_latency}, the block period $\BlockPeriod$ affects transactions block inclusion latency.
A longer block period implies higher latency.
On the other hand, a shorter period results in a higher number of blocks created per time interval (since a block for each block period is created). 
Since each block has a fixed size header, the larger the number of blocks created, the higher the space occupied by block headers compared to transactions in the blockchain, that is, the higher the space overhead.

\begin{figure}[t]\centering
\includegraphics[width=0.6\linewidth]{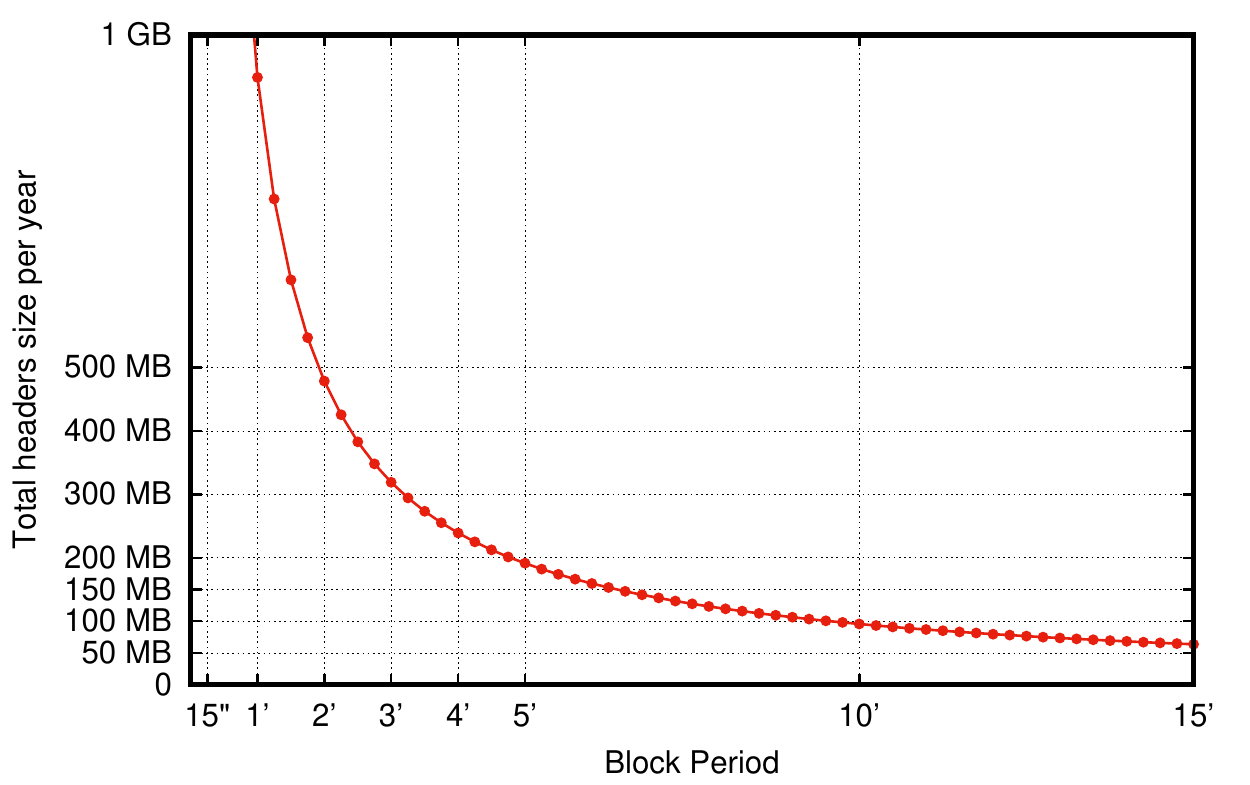}
\caption{Total headers size per year for different block periods $T$.}\label{fig:Space_Overhead}
\end{figure}

The headers size overhead, \ie the total size of block headers, at any time $t$ is:
\begin{equation}\label{eqn:space_overhead}
\HeadersOverhead(t) = \HeaderSize \cdot \frac{t}{\BlockPeriod}
\end{equation}
Note that this value only depends on the number of blocks in the chain at time $t$, and not on the number of transactions.
Figure~\ref{fig:Space_Overhead} shows the space overhead per year ($\HeaderSize = 1909$ bytes in our prototype implementation), that is, how much blockchain's space is taken up by block headers every year. 
For example, for $\BlockPeriod = 5$ minutes the space overhead is around $191$ MB per year. 
We find this value of $\BlockPeriod$ a good trade-off between transaction latency and block headers overhead for this particular application of the blockchain.

\subsection{Blockchain growth rate}\label{sec:growth_rate}

The blockchain can be seen as an append-only database. 
That is, its size cannot shrink over time.
If $I_{\TransactionSet}(t)$ is the set of transactions included in the blockchain at time $t$, then the blockchain total size at time $t$ is:
\[
\size_{\mathrm{bc}}(t) = \GenesisBlockSize + \operatorname{overhead}_{\mathrm{bc}}(t) +  \sum\limits_{\Transaction \in I_{\TransactionSet}(t)} \size(\Transaction)
\]
where $\GenesisBlockSize$ is the size of the genesis block.
Therefore, the growth rate over a time interval $[t_1, t_2]$ is $\size_{\mathrm{bc}}(t_2) - \size_{\mathrm{bc}}(t_1)$, that is:
\begin{equation}\label{eqn:growth_rate}
\GrowthRate(t_1;t_2) = \HeaderSize \cdot \frac{t_2 - t_1}{\BlockPeriod} + \sum\limits_{\Transaction \in I_{\TransactionSet}(t_1;t_2)}\size(\Transaction)
\end{equation}
where $I_{\TransactionSet}(t_1;t_2) = I_{\TransactionSet}(t_2) \setminus I_{\TransactionSet}(t_1)$.

Obviously, how fast a blockchain grows over time depends mainly on the transaction rate. 
Another factor that affects the growth rate is the block period~$\BlockPeriod$.
As already shown in section~\ref{sec:space_overhead}, this parameter affects the headers overhead and thus the first term of equation~\ref{eqn:growth_rate}.
The block gas limit parameter $\GasLimit$ may also affect the growth rate, as, if not properly dimensioned, it may increase latency spreading the incoming transaction rate over a larger time period, thus, decreasing the growth rate (i.e., $\GasLimit$ would affect the number and type of transactions included in $I_{\TransactionSet}(t_1;t_2)$ and thus the second term of equation~\ref{eqn:growth_rate}).
However, the analysis detailed in section~\ref{sec:tx_latency}, should allow to set the value of $\GasLimit$ so as to bound transaction latency.
In practice, $\GasLimit$ should be set greater than the average gas rate to avoid an ever increasing latency.
In this conditions, if the growth rate is computed over a large enough interval of time (to hide the effects of potential peak gas rate periods), the block gas limit parameter should not affect the growth rate significantly (that is, if properly set, $\GasLimit$ should not affect $I_{\TransactionSet}(t_1;t_2)$).
Otherwise, $\GasLimit$ should be set to a larger value.

\begin{table}[t]
\centering
\caption{Growth rate for different classes of workloads ($n$ \texttt{CreateEvidence(1024)}, $n$ \texttt{RemoveEvidence()}, $10n$ \texttt{Transfer()} per year).}
\label{tab:growth_rate}
\begin{tabular}{|l|l|l|c|}
\hline
 Workload & $\GrowthRate - \HeadersOverhead$ & $\GrowthRate$ with $\BlockPeriod = 5'$ & $\HeadersOverhead / \GrowthRate$ (\%) \\
\hline
\hline
$n = 10000$    &  $29.7$ MB/year &  $221.08$ MB/year & $86.56\%$ \\
\hline
$n = 100000$    &  $297$ MB/year &  $488.45$ MB/year & $39.18\%$ \\
\hline
$n = 1000000$    &  $2.9$ GB/year &  $3.09$ GB/year & $6.05\%$ \\
\hline
\end{tabular}
\end{table}

By using equation~\ref{eqn:growth_rate} we computed the annual growth rate for different classes of workloads. 
Since we were not able to find any publicly available statistics about evidence collection and transfer, we considered different classes of synthetic workloads with $n$ new evidence creations and removals and $10n$ transfers per year.
The results of this analysis are reported in Table~\ref{tab:growth_rate}.
The second column of Table~\ref{tab:growth_rate} reports the annual growth rate without considering the headers size overhead, while the third one includes the overhead term computed for $\BlockPeriod = 5'$. 
Finally, the fourth column shows the overhead percentages.
Even in presence of a very large number of evidence collection (1 million per year) and transfers (10 millions per year) the growth rate is around $3$ GB per year, which seems acceptable given the capacities of todays storage devices.



\section{Conclusion}\label{sec:conclusion}

This paper presented B-CoC, a blockchain-based architecture to dematerialise the CoC process in digital forensics.
We also provided a prototype of the B-CoC architecture based on the Geth implementation of Ethereum nodes. 
Based on the performance evaluation, B-CoC showed to be an effective support for the CoC process as it is able to sustain realistic workload with an acceptable overhead in terms of memory used to store the chain.

The current implementation assumes that the set of validators node is fixed and that validators are available to sacrifice their privacy when participating in the consensus process. As a future work, we are investigating how it is possible to manage a dynamic set of validators and most important we are studying alternatives that allow to increase the level of privacy for validators not altering other dependability and security attributes.


%
%
%
\bibliographystyle{splncs04}
\bibliography{bibliography}


\newpage

\begin{appendices}
\renewcommand{\thesection}{\Alph{section}}
\newcommand{\GasLimitLB}{\GasLimit_{\mathrm{L}}}
\newcommand{\GasLimitUB}{\GasLimit_{\mathrm{U}}}
\newcommand{\GasLimitLBavg}{\GasLimitLB^{\mathrm{avg}}}

\noindent {\Large{\bf Appendix}}

\section{Smart Contract Code}\label{sec:smart-contract}

\begin{lstlisting}[caption=Smart contract code.,label=lst:smart-contract,language=Solidity,escapechar=@]
pragma solidity ^0.4.22;
contract ChainOfCustody {
	struct Evidence {			@\label{line:struct-start}@
		bytes32 ID; 
		address owner;
		address creator;
		string description;
		address[] taddr; 
		uint[] ttime;
	}							@\label{line:struct-end}@
	
	mapping(bytes32 => Evidence) private evidences;	@\label{line:map}@
	
	modifier OnlyOwner(bytes32 ID) {
		require(msg.sender == evidences[ID].owner);
		_;
	}
	modifier OnlyCreator(bytes32 ID) {
		require(msg.sender == evidences[ID].creator);
		_;
	}
	modifier EvidenceExists(bytes32 ID, bool mustExist) {
		bool exists = evidences[ID].ID != 0x0;
		if (mustExist)
			require(ID != 0x0 && exists);
		else
			require(!exists);
		_;
	}
	function CreateEvidence (bytes32 ID, string description) 
			public EvidenceExists(ID, false) { @\label{line:create-evidence}@
		evidences[ID].ID = ID;
		evidences[ID].owner = msg.sender;
		evidences[ID].creator = msg.sender;
		evidences[ID].description = description;
		evidences[ID].taddr.push(msg.sender);
		evidences[ID].ttime.push(now);
	}
	function Transfer(bytes32 ID, address newowner) 
			public OnlyOwner(ID) EvidenceExists(ID, true) { @\label{line:transfer}@
		evidences[ID].owner = newowner;
		evidences[ID].taddr.push(newowner);
		evidences[ID].ttime.push(now);
	}
	function RemoveEvidence(bytes32 ID) 
			public OnlyCreator(ID) EvidenceExists(ID, true) { @\label{line:remove-evidence}@
		delete evidences[ID];
	}
	function GetEvidence(bytes32 ID) 
			view public returns(bytes32, address, address, 
				string, address [], uint []) { @\label{line:get-evidence}@
		return(evidences[ID].ID, evidences[ID].owner, 
			evidences[ID].creator, evidences[ID].description, 
			evidences[ID].taddr, evidences[ID].ttime);
	}
}

\end{lstlisting}

%

\newpage

\section{Discussion on the configuration of the parameters}\label{sec:discussion-params}

Section~\ref{sec:evaluation} discusses how the parameters of \arch affect its performance with respect to different aspects, namely the transaction latency, the block headers overhead and the blockchain growth rate.
Here we give a comprehensive discussion on how to set \arch parameters.

\subsection{Setting the block period $\BlockPeriod$}
The block period $\BlockPeriod$ affects transactions block inclusion latency (see section~\ref{sec:block_incl_latency}) and the block headers overhead (section~\ref{sec:space_overhead}), that ultimately affects the blockchain growth rate.
As already discussed in section~\ref{sec:space_overhead}, a shorter block period results in a lower maximum block inclusion latency, but also in a higher block header overhead.
To find the best trade-off one can use equation~\ref{eqn:space_overhead}.
For example, with our prototype implementation of \arch we consider $\BlockPeriod = 5'$ to be a good trade-off between latency and block header overhead.
Indeed, any further increase of $\BlockPeriod$ would result in a small improvement in terms of overhead reduction compared to the increase of latency (as shown in Figure~\ref{fig:Space_Overhead}).

\subsection{Setting the block gas limit $\GasLimit$}

The block gas limit affects both term of the transaction latency.
In particular in section~\ref{sec:tx_latency}, we describe an analysis that allows to derive a lower bound $\GasLimitLB$ for $\GasLimit$, to limit block inclusion latency and an upper bound $\GasLimitUB$ to limit consensus latency.

When $\GasLimitLB \leq \GasLimitUB$ it is safe to set $G$ equal to any value in $[\GasLimitLB, \GasLimitUB]$ to obtain a maximum block inclusion latency bound by $\BlockPeriod$ and the desired maximum consensus latency.
On the other hand, if $\GasLimitLB > \GasLimitUB$, it is not possible to have both terms of transaction latencies bounded by the desired values. 
In such case, one should set $\GasLimit$ to the best trade-off between block inclusion latency and consensus latency.
A good strategy may be to discard the lower bound $\GasLimitLB$ in favor of a new lower bound $\GasLimitLBavg$ which is set to the average gas rate rather than the maximum gas rate.
Setting $\GasLimit = \GasLimitLBavg$ would result in block inclusion latency bounded by $\BlockPeriod$ on average, with possible periods of increasing latencies, \eg during peak loads.
In this case, if $\GasLimitLBavg \leq \GasLimitUB$ one should set $\GasLimit = \GasLimitUB$, otherwise $\GasLimit = \GasLimitLBavg$.
Indeed, setting a value of $\GasLimit$ less than the average gas rate would result in ever increasing transaction latencies.

\end{appendices}

\end{document}